\documentstyle[aps,prl,epsfig]{revtex}
\newcommand{\bm}[1]{\mbox{\boldmath $#1$}}
\newcommand{\mean}[1]{\left< #1 \right>}

\title{Simple stochastic models showing strong anomalous diffusion}
\author
{K.H.~Andersen$^{1}$, P.~Castiglione$^{1,2}$, A.~Mazzino$^{3}$
and A.~Vulpiani$^{1}$}
\address{ $^{1}$  Dipartimento di Fisica, and INFM,
Universit\`a ``La Sapienza'',\\ P.le A. Moro 2, 00185 Roma, Italy.}
\address{ $^{2}$ Laboratoire de Physique Statistique,
Ecole Normale Sup\'erieure,
 24 rue Lhomond, 75231 Paris Cedex 05, France.}
\address{ $^3$ INFM--Dipartimento di Fisica, Universit\`a
di Genova, I--16146 Genova, Italy}
\begin{document}
\maketitle
\date{}

\begin{abstract}
  We show that {\it strong} anomalous diffusion, i.e.
  $\mean{|x(t)|^q} \sim t^{q \nu(q)}$ where $q \nu(q)$ is a nonlinear
  function of $q$, is a generic phenomenon within a class of
  generalized continuous-time random walks. For such class of systems
  it is possible to compute analytically $\nu(2 n)$ where $n$ is an
  integer number.  The presence of strong anomalous diffusion implies
  that the data collapse of the probability density function
  $P(x,t)=t^{-\nu}F(x/t^{\nu})$ cannot hold, a part (sometimes) in the
  limit of very small $x/t^{\nu}$, now $\nu=\lim_{q \to 0} \nu(q) $.
  Moreover the comparison with previous numerical results shows that
  the shape of $F(x/t^{\nu})$ is not universal, i.e., one can have
  systems with the same $\nu$ but different $F$.

\end{abstract}

PACS number(s): 05.45.+b, 05.60.+w;

\section{Introduction}
Anomalous diffusion, i.e., when the scaling of the moments of the
position $x(t)$ is $\mean{x^2(t)} \sim t^{2 \nu}$ with $\nu > 1/2$,
has been observed in a rather wide class of dynamical systems, e.g.,
intermittent maps \cite{1a,1b}, $2D$ symplectic maps \cite{2a,2b,2c}
and random velocity field \cite{3a,3b,3c} as well in $2D$
time-dependent flow \cite{4} and Hamiltonian systems (e.g.~the
egg-crate potential) \cite{5,6}.  In highly nontrivial systems, as
those described in \cite{4,5}, the existence of anomalous diffusion
has been established only numerically.  On the other hand, for random
shears it is possible to give an analytical criterium both for the
existence of anomalous diffusion and for the computation of $\nu$
\cite{multis}.  As far as we know, the simplest nontrivial system
showing anomalous diffusion is the continuous-time random walk (CTRW),
sometimes also called L\'evy walk.  The CTRW is entirely specified by
the probability density function (pdf) $\psi(r,\tau)$ to move a
distance $r$ in a time $\tau$ in a single motion event.  Let us
assume, as in \cite{7a,7b,7c},
\begin{equation}
  \psi(r,\tau)=P(\tau) \, P(r \mid \tau),
\label{eq1}
\end{equation}
where $P(\tau)$ is the pdf of having a flight of duration $\tau$ and
$P(r \mid \tau)$ is the conditional pdf of a displacement $r$ given
the flight time $\tau$. The cases corresponding to $P(\tau)\sim
\tau^{-g}$ and $P(r \mid \tau)=\delta(\mid r \mid - \tau^{\alpha})/2$
can be treated analytically \cite{7a,7b,7c}.

If the scaling of all moments can be described by just one exponent,
i.e. $\mean{x^{2n}(t)} \sim t^{2 n \nu}$, a collapse of the pdf's at
different times is obtained exploiting the rescaling \cite{4}
\begin{equation}
P(x,t)=t^{-\nu}F(x/t^{\nu})~.
\label{eq2}
\end{equation}
Then it also becomes clear that the value of $\nu$, in general, does not
completely characterize the statistical properties of the diffusion
process, as the function $F(\xi)$ needs to be specified.

In many cases, the use of just one exponent is not enough to describe
all the moments, i.e., we have the so-called {\it strong} anomalous
diffusion \cite{4,5}. For which $\mean{|x(t)|^q} \sim t^{q \nu(q)}$,
where $q \nu(q)$ is a nonlinear function of $q$. The existence of a
non-unique scaling exponent implies the failure of the data collapse
for the pdf in the form given by Eq.~\ref{eq2}. The best known case of
a process showing strong anomalous diffusion is the advection of a
passive scalar by a turbulent velocity field. In many cases \cite{4}
it has been observed that the function $q \nu(q)$ is piecewise linear,
i.e.,
\begin{equation}
q \, \nu(q) \simeq \left\{ \begin{array}{ll}
 \nu_1 q   & \quad  q < q_c \\
 q-c  & \quad q > q_c.
\end{array}
\right.
\label{eq3}
\end{equation}
This is basically due to the existence of two mechanisms: a {\it weak}
(i.e., with a unique exponent $\nu_1 > 1/2$) anomalous diffusion for
the typical events, and a ballistic transport for the rare excursions
(i.e., excursions much larger than $x_{typ}(t) \equiv \exp\mean{\ln
x(t)}$). The behavior (\ref{eq3}) suggests the validity of the data
collapse (\ref{eq2}) for the pdf core, i.e.  $x/t^{\nu_1}$ not too
large, and two peaks at $\mid x \mid \sim t$ ,i.e.,~the footprint of
ballistic events. The cases where strong anomalous diffusion for which
(\ref{eq3}) holds are relatively simple and surely different from the
cases of the relative dispersion in the fully developed turbulence
\cite{8}.

By using only elementary techniques, in this paper we show that the
bi-linear behavior for the scaling of the moments (\ref{eq3}), which
is present in the special case of the CTRW commonly found in the
literature, i.e.~$P(r \mid \tau)=\delta(\mid r \mid - \tau^{\alpha})/2$,
does not hold in the general case. To show this point, we shall
consider a generalized CTRW of the form:
\begin{equation}
P(r \sim \tau^{1+h} \mid \tau)\sim \tau^{-S(h)}
\label{eq4}
\end{equation}
The inspiration for this choice comes from the multifractal
description of turbulence \cite{Libro}.

The paper is organized as follows. In section \ref{sec:standard} we
present the ``standard'' CTRW model. We then present a simple method
to find the scaling for the even order moments (section
\ref{sec:scaling}). In section \ref{sec:general} the model is
generalized, and the same method is again applied to find the scaling.
Numerical analysis to corroborate the analytical results of the
general model are presented in Section \ref{sec:numerics} together
with some discussions related to the shape of $P(x,t)$. Discussions and
conclusions can finally be found in Section \ref{sec:discussion}.


\section{CTRW models}
\label{sec:general_model}
Anomalous diffusion occurs when some, or all, of the hypothesis of the
central limit theorem break down. More specifically, the system has to
violate at least one of the two following conditions:
\begin{enumerate}
\item Finite variance of the velocity.
\item Fast enough decay of the auto-correlation function of the
Lagrangian velocities.
\end{enumerate}
The paradigmatic model for anomalous diffusion, namely the L\'evy
flights \cite{14} violate the first condition. In the one dimensional
case a L\'evy flight corresponds to the evolution in discrete time
\begin{equation}
x(t_{i+1})=x(t_{i})+v_{i}\Delta t
\label{LF}
\end{equation}
of the particle position $x$ with $t_{i+1}=t_{i}+\Delta t$ and $v_{i}$
being independent stochastic variables identically distributed
according to a L\'evy-stable distribution such that
\begin{equation}
  P(v)\sim v^{-g} \,\, {\rm for\;\; large}\;\; v
\end{equation}
where $1< g \leq 3$.
It is easy to show that $\langle x^{2}\rangle =\infty $ for $g <3$ and
that this stochastic process shows anomalous diffusion being
$x_{typ} \sim t^{1/(g-1)} > t^{1/2}$.

\subsection{The ``standard'' CTRW model}
\label{sec:standard}
The existence of an infinite variance is not very pleasing from a
physical point of view. This has lead to the introduction of the CTRW
(also called L\'evy walks). The idea is to relax the condition of a
fixed, discrete time step in such a way that the process still has
anomalous diffusion, but finite variance of the velocity.

Firstly, we introduce the particle trajectory
\begin{equation}
x(t)=x(t-\tau_i)+ v_i \tau_i
\label{eq2.1}
\end{equation}
where $x(t)$ denotes the position of the particle at time $t$.  During
the random intervals $\tau_i$, the particles move a distance $r_i$
with constant random velocity $v_i$ independent of $\tau_i$. After
each interval they choose new random values for $\tau_i$ and $v_i$.

The relevant quantity to characterize the motion of the particle is
the pdf $\psi(r,\tau)$ of having a displacement $r$ in time $\tau$ in
a single motion event. This pdf is chosen in the form (\ref{eq1}).  In
the simple case where $v_i=\pm v$ we have
\begin{equation}
  P(r \mid \tau)=\frac{1}{2}\delta(\mid r \mid - v \tau)~.
  \label{eq2.3}
\end{equation}
Taking
\begin{equation}
  P(\tau) \sim \tau^{-g} \,\,\,\, {\rm for\;\; large}\;\; \tau
  \label{eq2.4}
\end{equation}
we can determine the pdf $P(x,t)$ to be in $x$ at time $t$
\cite{Proceed1,Proceed2}.
Actually, by introducing the probability density $\Psi(x,t)$ to pass
at location $x$ at time $t$ in a single motion event (and not
necessarily to stop at $x$)
\begin{equation}
  \Psi(x,t)=P(x \mid t) \int_t^{\infty} d\tau \int_{\mid x \mid}^{\infty}
  d r \; \psi(r,\tau)=
  \frac{1}{2}\delta(\mid x \mid - v t) \int_t^{\infty} d\tau \;P(\tau)
  \label{eq2.3b}
\end{equation}
the pdf $P(x,t)$ can be written in the following way
\begin{equation}
  P(x,t)=\Psi(x,t)+\int_{-\infty}^{\infty} dx' \int_0^{t}d\tau \;
\psi(x',\tau)
  \Psi(x-x',t-\tau) +...
  \label{eq2.3c}
\end{equation}
the first term denotes the probability density to reach the position
$x$ at time $t$ in a single motion event, the second term is the
probability density to reach $x$ at time $t$ with one stop in $x'$ and
so on to include all the combinations of motion events.  In the
Fourier-Laplace space $(x \rightarrow k, t \rightarrow u)$ the series
in Eq.~(\ref{eq2.3c}) assumes the closed form
\begin{equation}
  \hat{P}(k,u)=\frac{\hat{\Psi}(k,u)}{1-\hat{\psi}(k,u)}
  \label{eq2.3d}
\end{equation}
and the behavior of $\langle x(t)^2\rangle$ can be calculated analytically
by using the relation
$$\langle x(t)^2\rangle=-{\cal L}^{-1} \left[ \frac{\partial^{2}}
{ \partial^2 k}\hat{P}(k,u) \mid_{k=0} \right],$$ where
${\cal L}^{-1}$ denotes the inverse Laplace transform $(u \rightarrow t)$
\cite{Proceed1,Proceed2}.

The results is that
\begin{equation}
  \langle x(t)^2 \rangle \sim \left\{
    \begin{array}{ccccl}
      t^2             & 1 &< g < & 2 & {\mathrm ballistic\ motion}\\
      t^{4-g}         & 2 &< g < & 3 & {\mathrm anomalous\ diffusion}\\
      t               & 3 &< g  &   &  {\mathrm normal\ diffusion}.
    \end{array}
  \right.
  \label{eq2.6}
\end{equation}
Thus, enhanced anomalous diffusion occurs for $2\!<\!g\!<\!3$. For the
moments of small order, which describe the core of the pdf, it has
been shown that the asymptotic behavior gives $\mean{|x(t)|^q} \sim
t^{q\nu}$ with $\nu = 1/(g-1)$ \cite{Proceed1,Proceed2}. Thus the core
of the pdf can be scaled as in (\ref{eq2}) using $\nu$. The ballistic
motions which are responsible for the different scaling of higher
order moments, show up as wings on the pdf, which does not scale using
$\nu$.

The previous approach can be generalized \cite{7d} to the case where
\begin{equation}
 P(r \mid t)=\frac{1}{2}\delta(\mid r \mid - \tau^{\alpha}),\;\;\;v=\pm
 \tau^{\alpha-1},\;\;\;
 P(\tau)\sim \tau^{-g} \;\;{\rm with}\;\; g > 1.
 \label{eq2.15a}
\end{equation}
In addition one can treat more complicated situations \cite{5,7c} by
considering that the particle can move ballistically but it can be
also trapped in some structures as vortices \cite{SoloGollu} or
chaotic islands (standard map or ``egg-crate'' potential)
\cite{Proceed2,Geisel88}.

\subsection{Finding the scaling of the moments}
\label{sec:scaling}
The usual method used in \cite{7d,Proceed1,Proceed2} to find the
scaling of the moments in the simple CTRW model is not elementary. We
now present an alternative easier way to calculate the displacement
moments $\mean{x^{q}(t)}$ and thus to characterize the anomalous
diffusion. 

Let us consider a particle moving ballistically with
velocity $v_i$ during the interval times $\tau_i$. The velocities
$v_i$ are identically distributed, independent, random variables
assuming the values $\pm 1$ alternatingly. The intervals times
$\tau_i$ are identically distributed, independent, random variables
and assuming the value $\tau$ with probability 
\begin{equation}
  P(\tau)\sim \tau^{-g} \;\;{\rm with}\;\; g > 1 \;\;\; {\rm and} \;\;\;
\tau\in[t_{min},T]
\label{eq2.15bis}
\end{equation}
where $t_{min}$ and $T$ are the lowest and highest cutoffs, respectively.
The reason for which we need to introduce such cutoffs will be clear
later. The particle position at the time $t$ can be written
as
\begin{equation}
 x(t) = \sum_{i=1}^n v_i \tau_i + \,v_{n+1} \epsilon_{n+1} \;\;{\rm
   with} \;\; \epsilon_{n+1}= t-\sum_{i=1}^n \tau_i
\label{eq2.16}
\end{equation}
$n$ being the (random) integer value for which $t_n\leq t$ and
$t_{n+1} > t$. The total time $t$ can be rewritten as $t=\sum_{i=1}^n
\tau_i+\epsilon_{n+1}$. Denoting by $N\equiv\langle n \rangle$ the
average value of the number of time steps necessary to reach the time
$t$, one has for large times  $t\simeq N\langle\tau\rangle$ and
thus $x(t) = \sum_{i=1}^n v_i\tau_i$.

 From simple considerations related to the symmetry of the velocity
 pdf under the transformation $v\mapsto -v$, it immediately follows
 that the odd-order moments $\mean{x(t)^q}$ are trivially zero.
 Conversely, even-order moments are nonzero and can be
 evaluated exploiting the following properties: $\mean{v_i\tau_j}=0$,
 $\mean{v_i v_j}\propto\delta_{ij}, \mean{\tau_i \tau_j}\propto
 \delta_{ij}$.  For times large enough, the mean squared displacement
 thus reads:
\begin{equation}
 \mean{x(t)^2} = \mean{\sum_{i=1}^{n}\sum_{j=1}^{n} (v_i \,\tau_i)\, (v_j
\,\tau_j)} =  N \mean{(v\, \tau)^2}.
\label{eq2.19}
\end{equation}
Similarly, for the fourth and sixth order moments the limit of large
times yields
\begin{eqnarray}
\mean{x^4(t)} &=& N\; \mean{(v \tau)^4} + 3\; N^2\; \mean{( v \tau)^2}^2
\nonumber\\
\mean{x^6(t)} &=& N\; \mean{(v \tau)^6} + 15\; N^2\; \mean{( v
\tau)^2}\mean{( v \tau)^4}+
\nonumber \\
& & 15\; N^3\;\mean{( v \tau)^2}^3.
\label{eq2.21}
\end{eqnarray}
and so on.

Our attention being focused on the behavior of $\mean{x(t)^q}$ as a
function of $t$, we make the substitution $N=t/\mean{\tau}$ in previous
expressions (\ref{eq2.19}) and (\ref{eq2.21}) and again exploit the facts that
$v_i$ and $\tau_i$ are uncorrelated and $\mean{v_i^2}=1$ to get:
\begin{eqnarray}
\mean{x^2(t)} &=& \frac{t}{\mean{\tau}} \mean{\tau^2} \nonumber \\
\mean{x^4(t)} &=& \frac{t}{\mean{\tau}} \mean{\tau^4} + 3 \left
(\frac{t}{\mean{\tau}}\right )^2
\mean{ \tau^2}^2 \nonumber\\
\mean{x^6(t)} &=& \frac{t}{\mean{\tau}} \mean{\tau^6} + 15
\left (\frac{t}{\mean{\tau}}\right )^2\;
\mean{\tau^2}\mean{\tau^4} + \nonumber\\
& & 15\left (\frac{t}{\mean{\tau}}\right )^3\mean{\tau^2}^3.
\label{eq2.22}
\end{eqnarray}
For times $t \gg T$ the leading term for $\mean{x^{2 n}(t)}$ is that
one proportional to $t^n$, therefore one has:
\begin{equation}
\mean{x(t)^{2n}} \propto
   t^{n}\left( \frac{\mean{\tau^2}}{\mean{\tau}}\right )^{n} \quad
   {\mathrm where\ }n{\mathrm \ is\ integer.}
\end{equation}
which is just ordinary diffusive behaviour.  Diffusive behavior in
such limit is actually expected from general considerations. Indeed,
when $t \gg T$ the particle position at the time $t$ can be rearranged
in the form of a sum of almost independent displacements. If the
number of the latter is large enough, central limit arguments apply,
with the immediate consequence is that particles undergo diffusive
motion. Such result can also be rigorously proved exploiting multiscale
perturbative expansions in the (small) parameter $T/t$ as done, e.g.,
in Ref.~\cite{multis}.

In the opposite regime, where $t \ll T$, the system is strongly
correlated, central limit arguments does not apply, and the final
result is that non-diffusive (i.e. anomalous) regimes can occur where
$\mean{x^q(t)}\sim t^{q \nu(q)}$ with $\nu(q)\neq 1/2$.

The possible emergence of anomalous behaviors in the limit
$t\to\infty$ can be investigated by looking at the dependence of
moments on the cutoff $T$. For times shorter than $T$ they behave as
$\mean{x^q(t)} \sim t^{q\nu(q)}$, but around $t\sim T$ the moments
have a crossover to diffusive behaviour, $\mean{x^q(t)}
\sim(t/\mean{\tau})^{q/2}\mean{\tau^2}^{q/2}$.

By matching the two different regimes at $t = T$, and using the
results $\mean{\tau^q}\sim T^{-g+q+1}$ for $-g+q+1 > 0$ and
$\mean{\tau^q}=O(1)$ for $-g+q+1 < 0$, the following expressions for
$q \nu(q)$ are found as a function of the exponent $g$:
\begin{equation}
\begin{array}{llll}
g \in (1,2] \;\;\; & q \nu(q)=q\;\;\; & q=2,4,6,... &\\
& & &\\
g \in (2,3] \;\;\; & q \nu(q)=q+2-g\;\;\; & q=2,4,6,...& \\
& & &\\
g \in (3,4] \;\;\; & q \nu(q)=q/2  \;\;\; & q=2 & \\
                   & q \nu(q)=q+2-g\;\;\; & q=4,6,8,...&
\end{array}
\label{risulato1}
\end{equation}
and so on for higher values of $g$.

From (\ref{risulato1}) it follows that the anomalous diffusion phase
takes place for higher and higher moments as $g$ increases. For $q$
large enough one has $q\nu(q) = q + {\mathrm constant}$ and noting that
$2\nu(2) \ne 2$ one can conclude that $\nu(q)$ can not be constant and
therefore a strong anomalous diffusion is present.

The matching argument had been used in Ref.~\cite{multis} in the
context of the multiscale method for the anomalous diffusion in random
shear flows \cite{3a}.

\subsection{Generalized CTRW model}
\label{sec:general}
We now present a generalization of the previous model showing a {\it
  strong} anomalous diffusion regime characterized by a non-piecewise
linear behavior as a function of the order $q$.

As in the previous case the particle moves ballistically with random
velocity $v_i$ during the random interval times
$\tau_i=t_{i+1}-t_{i}$. We assume that $\tau_i$ has the same pdf as in
(\ref{eq2.15bis}) but now the velocities $v_i$ assume the value $\pm
\tau_i^h$ where $h$ is a random positive variable conditioned on $\tau_i$.
Specifically, the conditional pdf $P(h \mid \tau)$ is
\begin{equation}
P(h \mid \tau) \propto \tau^{-S(h)}
\label{eq2.g.1}
\end{equation}
where $S(h)$ is a positive smooth concave function. This has the
effect of giving a larger variance to the velocity, the larger the
time $\tau_i$. The function $S(h)$ can be taylored to a special need,
i.e., to mimic the intermittent turbulent velocity field. Here we just
use a generic, simple function $S(h) = h^2/(2\sigma^2)$, as we are
mainly interested in the generic properties of the model.

The moments $\mean{(v \;\tau)^{2q}}$ can now be calculated:
\begin{eqnarray}
\mean{(v \;\tau)^{2q}} &=& \int_{0}^{+\infty} dh \int_{t_{min}}^{T}
d\tau \; (v \tau)^{2q} P(\tau) P(h \mid \tau) \sim \\
                  & & \int_{t_{min}}^{T} d\tau \;\tau^{-g+2q}
\int_{0}^{+\infty}
dh \; \tau^{2q\,h-S(h)}=\\
                  & & \int_{t_{min}}^{T} d\tau\; \tau^{-g+2q+y(2q)}\sim
T^{-g+2q+y(2q)+1}.
\label{sella}
\end{eqnarray}
where the integral $\int dh \; \tau^{2q \,h-S(h)}$ has been evaluated
exploiting the steepest descent method and we have defined
\begin{equation}
y(2q)\equiv\max_{h} [2q \,h-S(h)].
\label{sellino1}
\end{equation}
Now using the specific shape of $S(h)$, expression (\ref{sellino1})
takes the form:
\begin{equation}
y(2q)=q^2 y(2) \;\; {\rm with}\;\; y(2)=2 \;\sigma^2.
\label{sellino2}
\end{equation}
Exploiting the same matching arguments as in section \ref{sec:scaling},
the following expressions for the exponents $q \nu(q)$ are obtained:
\begin{equation}
\begin{array}{llll}
g \in (1,2] \;\;\; & q \nu(q)=\frac{q^2}{2}\sigma^2 +q\;\;\; &
q=2,4,6,...&\\
& & &\\
g \in (2,3+y(2)] \;\;\; & q \nu(q)=\frac{q^2}{2}\sigma^2
+q+2-g\;\;\; & q=2,4,6,...&\\
& & &\\
g \in (3+y(2),4+y(4)] \;\;\; & q \nu(q)=q/2  \;\;\; & q=2 &\\
                   & q \nu(q)=\frac{q^2}{2}\sigma^2+q+2-g\;\;\;
& q=4,6,8, ...&
\end{array}
\label{risulato2}
\end{equation}
and so on, for higher values of $g$.

We now see that in the general case of the CTRW, $q\nu(q)$ is not just
a bi-linear function, and the exact form depends upon the shape of
$S(h)$. The generalization to an arbitrary shape of $S(h)$ is 
straightforward.

As far as we know it is not simple to obtain the results in
Eq.(\ref{risulato1})
and Eq.(\ref{risulato2}) with the method discussed in section II A.

\section{Numerical simulations}
\label{sec:numerics}
We present results of numerical simulations of the general model
defined by Eqs.~(\ref{eq2.15bis}) and (\ref{eq2.g.1}). The goal of the
numerical simulations has been to verify the validity of the
theoretical expectations (\ref{risulato1}) and (\ref{risulato2}) and
to study the pdf of $x(t)$ in more detail.

In practice the pdf for the length of the jumps $P(\tau)$ has to be
supplemented with a lower cutoff. As it is the tail of the
distribution which governs the scaling of the moments of $x(t)$ the
results are independent upon how the cutoff is made. We used
a hard cutoff at $\tau = 1$, such that $P(\tau)\!=\!0$ for $\tau < 1$
and
\begin{equation}
  P(\tau) = (g-1) \tau^{-g} \quad {\mathrm for}\quad \tau>1~.
\end{equation}
The moments have been calculated as an ensemble average of many
realizations of the process for times $0<t<\tilde{T}$. For all the
simulations presented here, $\tilde{T}$ has been set to $10^6$.

\begin{figure}[htbp]
  \begin{center}
    \epsfig{file=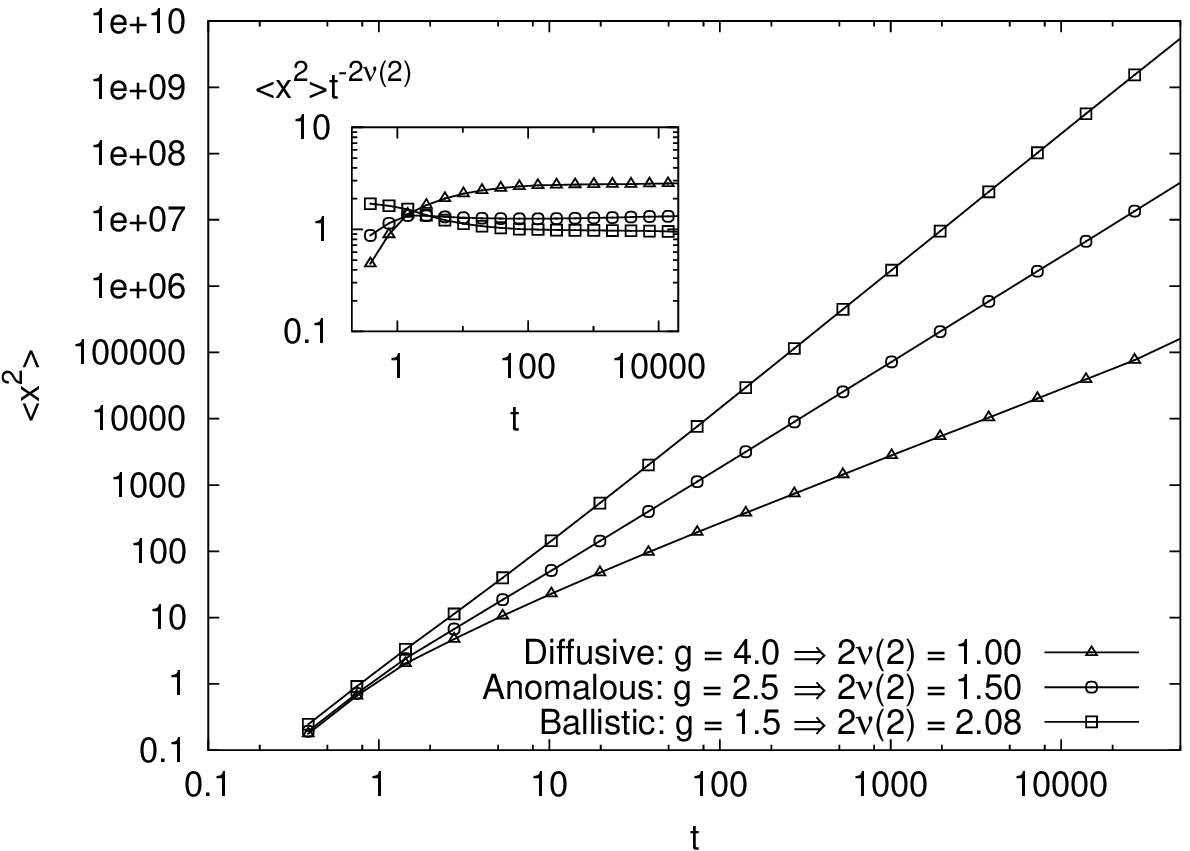}
    \caption{The scaling of the displacement $\left< x^2(t) \right>$
      for three different values of $g$, each corresponding to
      diffusive behavior ($g\!=\!4.0$), anomalous diffusion
      ($g\!=\!2.5$) and ballistic motion ($g\!=\!1.5$). The value of
      $\sigma$ is 0.2 for all the cases. In the inset the same is
      shown, but this time rescaled with the expected behavior
      $\left< x^2 \right> \propto t^{2\nu(2)}$, such that a scaling in
      accordance with the theoretical prediction corresponds to a
      horizontal line.}
    \label{fig:corr}
  \end{center}
\end{figure}

In Fig.~\ref{fig:corr} the scaling of the second moment $\left< x^2(t)
\right>$ with $t$ for different values of $g$ and $\sigma\!=\!0.2$ is
shown.  Examples have been chosen where diffusive, anomalous and
ballistic behavior is expected. After an initial ballistic motion for
short times, a transient towards scaling is taking place. For long
times, clean scaling with an exponent corresponding to
Eqs.~(\ref{risulato1}) is evident.

\begin{figure}[htbp]
  \begin{center}
    \epsfig{file=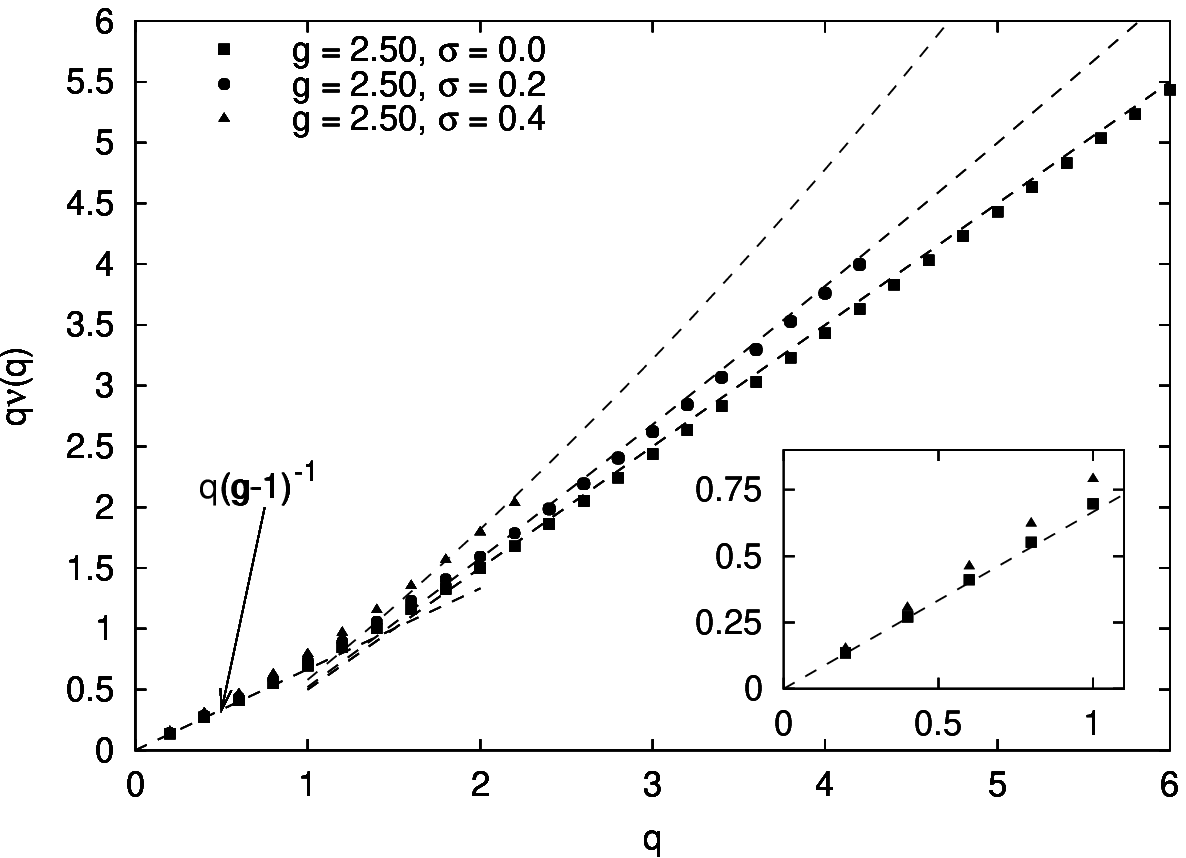}
    \caption{The scaling of the higher order moments for three
      examples, with $g = 2.50$ and $\sigma=0$, $0.2$ and $0.4$.
      The line originating from $q=0$ corresponds to $1/(g\!-\!1)$. The
      other curves corresponds to $q\nu(q)\!=\!q+2-g+(q\sigma)^2/2$.}
    \label{fig:multi}
  \end{center}
\end{figure}

To show how the introduction of the velocity field from section
\ref{sec:general} influences the moments, we have calculated the
higher order moments of $x(t)$ for $\sigma=0$,$0.2$ and $0.4$ for
$g=2.5$ (Fig.~\ref{fig:multi}). For the case with $\sigma\!=\!0$ we
see a clear bi-linear behavior as expected, with a cross over at $q_c
\approx 1.5$. For low order moments ($q<1$) the 
core of the pdf is important, and the behavior approches $1/(g-1)$.
As is seen in the inset, this behavior is only to be strictly valid in
the limit $q \rightarrow 0$. The scaling of the moments changes
smoothly into the prediction $\nu(q) = q+2-g$ for $q
\ge 2$. Using the fact that $q\nu(q)$ is a concave increasing function
and, for $\sigma=0$, the slope can not be larger than 1, one obtains
that the prediction given by Eq.~\ref{risulato1}, obtained only for
even order moments, surely is valid for any moment larger than 2. For
the generalized model ($\sigma>0$), the fluctuations become much
stronger, and it was not possible to get a clear convergence for
large orders.  However, for $q=2$ the prediction in
Eq.~\ref{risulato2} and the numerical results are in perfect
agreement.

\begin{figure}[htbp]
  \begin{center}
    \epsfig{file=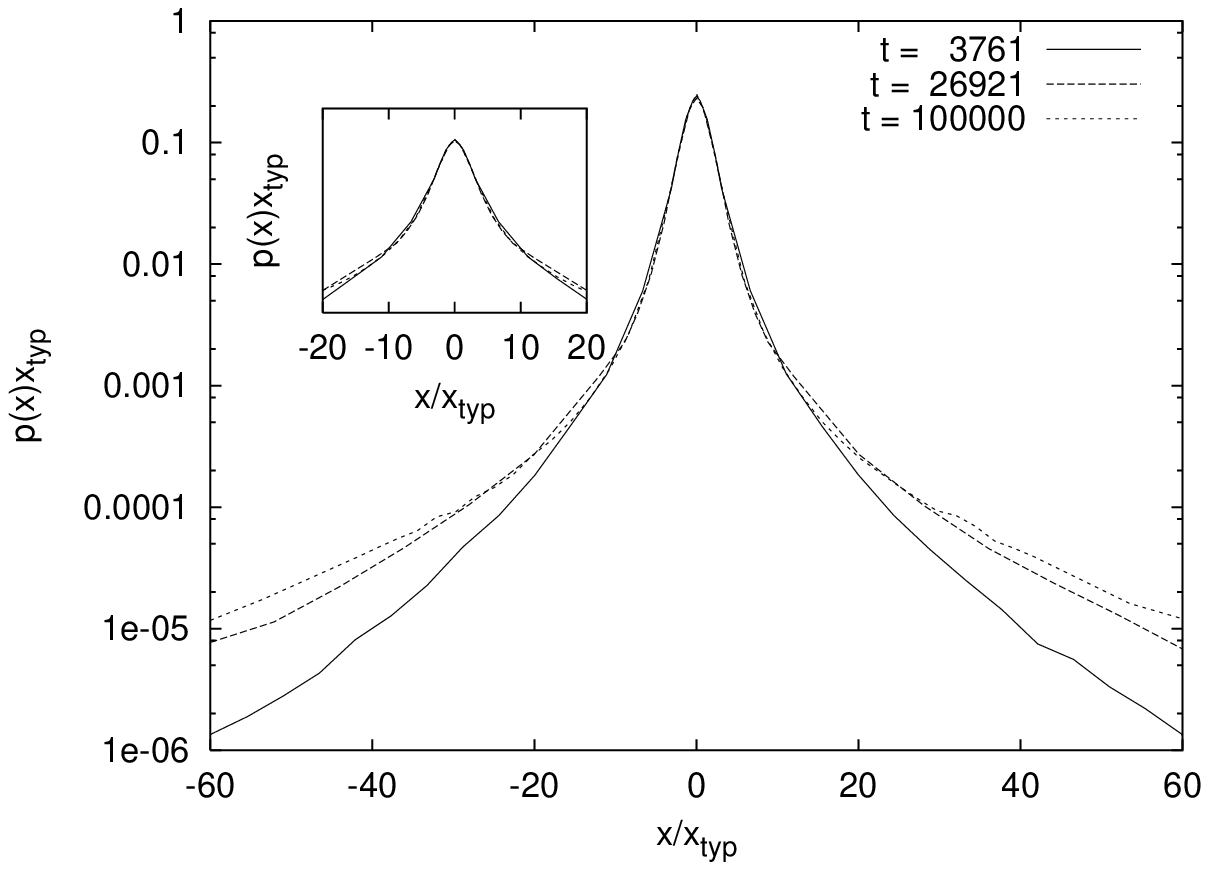}
    \caption{The pdf of $x(t)$ at three different
      times rescaled according to Eq.~(\ref{eq2}) for a situation with
      $g\!=\!2.5$ and $\sigma\!=\!0.2$ corresponding to the anomalous
      regime. The pdf have been scaled with the typical value,
      $x_{typ}(t)$.}
    \label{fig:pdf-compare}
  \end{center}
\end{figure}

The characterization of the process by the scaling exponents of the
moments can be seen as a way to probe the pdf $p(x,t)$ of
the process. Roughly speaking, the low order structure functions
characterize the core of the pdf, while higher orders characterize the
tails. In the case of ``ordinary'' anomalous diffusion, the pdfs can
be rescaled according to Eq.~\ref{eq2}, due to the fact that the whole
process can be characterized by just one scaling exponent.  However
for the bi-linear or the strong diffusive cases, no such
renormalization can be done, as more than one exponent is needed to
characterize the process. As all the cases seem to have the same
limiting behavior for $q \rightarrow 0$, it might be possible to make
a collapse of the core of the pdf. In Fig.~\ref{fig:pdf-compare} the
pdf for $g=2.5$ and $\sigma=0.2$ has been rescaled using the typical
value $x_{typ}(t) = \exp(\left< \ln(|x(t)|) \right>$. As expected the
core of the rescaled pdfs show a good overlap, while the tails
diverge. For higher values of $\sigma$ the area of the core with
overlap becomes smaller and smaller.

\section{Discussions and conclusions}
\label{sec:discussion}
By using only elementary techniques, we have shown in this paper that
the strong anomalous diffusion appears in CTRW. Our approach, in which
one computes the even moments $\mean{x(t)^q}$ for a system with a
cutoff $T$ on the pdf $P(\tau)$, and then the matching of the
behaviors at $t \ge T$ and $t \le T$, allows us to treat rather
general cases. This is a relevant advantage with respect to the
approach commonly used for CTRW which seems to us to be difficult to
apply for more general cases. In the case of generalized CTRW,
i.e. with Eq.~\ref{eq2.g.1}, one has a nontrivial (non bi-linear)
shape of $q\nu(q)$.  This implies that the pdf $P(x,t)$ cannot be
written in the form (\ref{eq2}). On the other hand we found that the
rescaling (\ref{eq2}), with $\nu = \nu(0)$, is valid in a limited
range of $x/x_{typ}$. It is rather natural to wonder if, at least in
this limited range, the function $F(\xi)$ is determined by the
exponent $\nu(0)$.  From previous works \cite{BGKPR90,BG90} and the
results discussed in section \ref{sec:numerics}, it seems to us that
one has a negative answer: $\nu$ does not determine the function
$F(\xi)$.

As the shape of the function describing the scaling of the moments
looks similar to what one finds for the relative dispersion of a
passive scalar by a turbulent velocity field, one might like to use
the the CTRW model to describe this process. There is however some
differences that has to be taken into consideration. In fully
developed turbulence, a flight of duration $\tau$ can be considered as
associated to an eddy of size $l\sim\tau^{3/2}$.  In a process of
relative diffusion, one typically has an increase of the relative
distance with the time. This implies that a flight of duration $\tau$
is preferentially followed by another flight with a larger duration
than the previous. The easiest way for a realistic description of
relative dispersion in fully developed turbulence should thus consist
in a further generalization of the CTRW where a Markov ingredient,
i.e.~a dependence of the flight duration at the time $t$ on the
duration at the previous time, is introduced. It seems to us that this
is a nontrivial task; a first attempt in this direction can be found
in the recent work \cite{SBK99}.


\section*{Acknowledgments}
We are particularly grateful to Paolo Muratore-Ginanneschi for very
stimulating discussions and suggestions and J.~Klafter for useful
correspondence. The authors thank M.~H.~Jensen for fruitful
discussions and all the CATS staff for the nice hospitality at Niels
Bohr Institute in Copenhagen.  A.M.~was partially supported by the
INFM PA GEPAIGG01.  P.C.~and A.V.~are partially supported by
INFM (Progetto Ricerca Avanzata--TURBO) and by MURST (program no.
9702265437). P.C.~and K.H.A.~ acknowledge the support of the European
Commission's TMR Programme, Contract no.ERBFMRXCT980175, ``Intermittency
in turbulent systems''. Simulations were performed at CINECA
(INFM Parallel Computing Initiative).

\end{document}